\documentclass[preprint]{revtex4}

\usepackage{dcolumn}
\usepackage{amsmath}

\usepackage{graphicx}
\usepackage{pslatex}

\def\eref#1{(\ref{#1})}
\def\d{{\rm d}}
\def\e{{\rm e}}
\def\etal{{\it{}et~al.}}
\def\av#1{\langle#1\rangle}
\def\set#1{\lbrace#1\rbrace}
\def\Li{\mathop{\rm Li}}

\begin{document}

\title{Exact solutions of epidemic models on networks}
\author{M. E. J. Newman}
\affiliation{Santa Fe Institute, 1399 Hyde Park Road, Santa Fe, NM 87501}
\date{January 23, 2002}

\begin{abstract}
  The study of social networks, and in particular the spread of disease on
  networks, has attracted considerable recent attention in the physics
  community.  In this paper, we show that a large class of standard
  epidemiological models, the so-called susceptible/infective/removed
  models, and many of their generalizations, can be solved exactly on a
  wide variety of networks.  Solutions are possible for cases with
  heterogeneous or correlated probabilities of transmission, cases
  incorporating vaccination, and cases in which the network has complex
  structure of various kinds.  We confirm the correctness of our solutions
  by comparison with computer simulations of epidemics propagating on the
  corresponding networks.
\end{abstract}

\maketitle

Networks of various kinds have been the subject of much recent research
within the physics community~\cite{Strogatz01,AB01}.  Social
networks~\cite{WS98,ASBS00,Liljeros01}, technological
networks~\cite{AJB99,FFF99}, and biological networks~\cite{Jeong00,FW00}
have all been examined and modeled in some detail.  Most work has focussed
on structural properties of the networks in question---patterns of
connection between people, computers, species, and so forth.  Structure,
however, while important, is in most cases only a prerequisite to answering
the question of real interest: what is the behavior of networked systems?
One area in which some progress has been made towards answering this
question is the study of the spread of disease.  Recent simulation studies
and approximate analytical treatments suggest that network structure can
play a crucial role in defining the nature of a disease
epidemic~\cite{MN00a,KA01,PV01a}.

In this paper, we show that the most fundamental standard model of disease
propagation, the SIR model, and a large set of its generalized forms, are
exactly solvable on a broad class of networks, including networks with
social or community structure of various kinds (e.g.,~networks in which
people are distinguished by different roles that they play).  Our solutions
provide exact criteria for deciding when an epidemic will occur, how many
people will be affected, and how the network structure or the transmission
properties of the disease could be modified in order to prevent the
epidemic.

The SIR model~\cite{Bailey75} is a model of disease propagation in which a
population is divided into three classes: susceptible~(S), meaning they are
free of the disease but can catch it, infective~(I), meaning they have the
disease and can pass it on to others, and removed~(R), meaning they have
recovered from the disease or died, and can not longer pass the disease on.
There is a fixed probability per unit time that an infective individual
will pass the disease to a susceptible individual with whom they have
contact, rendering that individual infective.  Individuals who contract the
disease remain infective for a certain time period before recovering (or
dying) and thereby losing their infectivity.

To turn this process into a complete model of disease spread we also need
to know the pattern of contacts between individuals.  In the standard
treatments, and indeed in most of mathematical epidemiology, researchers
use the so-called ``fully mixed'' approximation, in which it is assumed
that every individual has equal chance of contact with every other.  This
is an unrealistic assumption, but it has proven popular because it allows
one to write differential equations for the time evolution of the disease
that can be solved or numerically integrated to determine the course of an
epidemic.  More realistic versions of the model have also been studied in
which populations are divided into groups according to age or other
characteristics.  The models are still fully mixed within each group
however.  In the real world, the pattern of contacts between individuals is
far from fully mixed, forming a social network with well-defined structure.
Here, therefore, we abandon the fully mixed approximation and turn our
attention instead to models in which the social network is explicitly
represented.

In this paper we also abandon two other unrealistic assumptions of the
usual SIR model, the assumptions that all contacts between individuals
represent equal probability of disease transmission, and that all
individuals who catch the disease remain infective for the same amount of
time.  We will allow the probability per unit time $r_{ij}$ of transmission
from an infective individual~$i$ to a susceptible individual~$j$ to be
drawn from any arbitrary distribution~$P(r)$.  We will also allow the
time~$\tau$ for which individuals remain infective to be drawn from any
arbitrary distribution~$P(\tau)$.  These generalizations increase the range
(and realism) of models to which our solutions are applicable.

Consider then a network of initially susceptible individuals represented by
the vertices of a graph.  The edges of the graph represent connections
between individuals by which disease can be transmitted.  These connections
might represent, for example, periodic physical proximity---two people
working in the same building perhaps, or living in the same house.

The crucial observation that makes our solutions possible is that SIR
epidemic processes are equivalent to (generalized) bond percolation
processes on the corresponding network of individuals and contacts.  This
correspondence appears first to have been pointed out by Grassberger for
the case of the simple SIR model with fixed probabilities of infection and
times of infectiveness~\cite{Grassberger83}.  More recently, it has been
observed numerically that the correspondence extends also to the case of
variable probabilities and times~\cite{WSS01}.  In fact, it is
straightforward to show that the above generalized SIR process on a network
corresponds to bond percolation on the same network with uniform bond
occupation probability
\begin{equation}
T = 1 - \int_0^\infty \d r\>\d\tau \> P(r) P(\tau)\, \e^{-r\tau}.
\label{defst}
\end{equation}
The quantity $T$, which we call the transmissibility of the disease, lies
in the range $0\le T\le1$ and represents the average total probability that
a susceptible individual will catch the disease from an infective contact.
Our solutions for SIR models are derived by combining this mapping to
percolation with a generating function technique similar to that introduced
by Moore and Newman~\cite{MN00b}.

One of the most important results to come out of recent work on networks is
the finding that the degree distributions of many networks are highly
right-skewed.  (Recall that the ``degree'' of a vertex is the number of
other vertices to which it is connected.)  In other words, most vertices
have only a low degree, but there are a small number whose degree is very
high~\cite{AJB99,ASBS00,AB01}.  It is known that the presence of these
highly connected vertices can have a disproportionate effect on certain
properties of the network.  Recent work suggests that the same may be true
for disease propagation on networks~\cite{PV01a,LM01}, and so it will be
important that we incorporate non-trivial degree distributions in our
models.  As a first illustration of our method therefore, we look at a
simple class of unipartite graphs studied previously by a number of
authors~\cite{BC78,MR95,NSW01}, in which the degree distribution is
specified, but the graph is in other respects random.

Suppose that the probability of a randomly chosen vertex in our graph
having degree $k$ is $p_k$.  We define two generating
functions~\cite{NSW01}
\begin{equation}
G_0(x) = \sum_{k=0}^\infty p_k x^k,\qquad
G_1(x) = {1\over z} \sum_{k=0}^\infty k p_k x^{k-1},
\label{defsg0g1}
\end{equation}
where $z=G_0'(1)$ is the mean vertex degree in the network.  These two
functions generate respectively the probability distributions of the
degrees of randomly chosen vertices, and vertices at the ends of randomly
chosen edges.  Not all edges leading from a vertex will be occupied however
(i.e.,~result in transmission of the disease).  The distribution of the
number $m$ of occupied edges around a randomly chosen vertex is generated
by
\begin{eqnarray}
G_0(x;T) &=& \sum_{m=0}^\infty \sum_{k=m}^\infty p_k
           \biggl({k\atop m}\biggr) T^m (1-T)^{k-m} x^m\nonumber\\
         &=& \sum_{k=0}^\infty p_k \sum_{m=0}^k
             \biggl({k\atop m}\biggr) (xT)^m (1-T)^{k-m}
          =  \sum_{k=0}^\infty p_k (1-T+xT)^k\nonumber\\
         &=& G_0(1+(x-1)T).
\label{defsg0xt}
\end{eqnarray}
And similarly the number around the vertex at the end of a randomly chosen
edge is generated by $G_1(x;T) = G_1(1+(x-1)T)$.  Now the generating
function $H_1(x;T)$ for the total number of people infected as a result of
a single transmission along an edge in the network must satisfy a
self-consistency condition of the form~\cite{MN00b,NSW01}
\begin{equation}
H_1(x;T) = x G_1(H_1(x;T);T).
\label{bondh1}
\end{equation}
And the distribution of the number of people affected by an outbreak
starting with a single disease carrier is generated by
\begin{equation}
H_0(x;T) = x G_0(H_1(x;T);T).
\label{bondh0}
\end{equation}

The average size $\av{s}$ of a disease outbreak is then given by the
derivative of $H_0$ with respect to~$x$:
\begin{eqnarray}
\av{s} &=& H_0'(1;T) = 1 + G_0'(1;T) H_1'(1;T)\nonumber\\
       &=& 1 + {G_0'(1;T)\over1 - G_1'(1;T)}
        =  1 + {T G_0'(1)\over1 - T G_1'(1)},
\label{avs}
\end{eqnarray}
where we have made use of Eq.~\eref{bondh1} and the fact that all
generating functions are 1 at $x=1$ if the distributions that they generate
are properly normalized.  Eq.~\eref{avs} diverges when $T$ is equal to the
critical value $T_c=1/G_1'(1)$, and this point marks the onset of epidemic
behavior.  For transmissibilities below this epidemic threshold, $T<T_c$,
all outbreaks are finite in size, no matter how large the network, and the
probability of any given individual being affected by an outbreak is zero
in the limit of large graph size.  For $T>T_c$ there is always a finite
chance of infection.  The fraction of the population that is infected in an
epidemic outbreak can be derived by observing that above $T_c$,
Eq.~\eref{bondh0} generates the size distribution of outbreaks
\emph{excluding} epidemics~\cite{NSW01}, and hence the size $S$ of the
epidemic is given by the solution of
\begin{equation}
S = 1 - G_0(u;T),\qquad u = G_1(u;T).
\label{epidemic}
\end{equation}
Unfortunately, it is not usually possible to find a closed form solution to
this last equation, but it can be solved numerically by iteration from a
suitable starting value of~$u$.

Note that it is not the case, even above $T_c$, that all outbreaks give
rise to epidemics of the disease.  There are still finite outbreaks even in
the epidemic regime, and the probability of an outbreak becoming an
epidemic at a given $T$ is equal to~$S$.  While this appears very natural,
it stands nonetheless in contrast to the standard fully mixed models, for
which all outbreaks give rise to epidemics above the epidemic transition
point.

As an example of this first simple epidemic model, consider SIR disease
outbreaks taking place on networks having a degree distribution with the
truncated power-law form
\begin{equation}
p_k = \biggl\lbrace \begin{array}{ll}
            0                                        & \mbox{for $k=0$}\\
            C k^{-\alpha} \e^{-k/\kappa} \qquad\null & \mbox{for $k\ge1$.}
      \end{array}
\label{powerlaw}
\end{equation}
where $\alpha$ and $\kappa$ are constants, and $C$ is set by the
requirement that the distribution be normalized.  This distribution is seen
in a number of networks in the real world~\cite{ASBS00,Newman01a}, and
includes both pure power-law and pure exponential distributions as special
cases.

Substituting Eq.~\eref{powerlaw} into Eq.~\eref{defsg0g1}, we then find
that our disease has an epidemic transition at
\begin{equation}
T_c = {\Li_{\alpha-1}(\e^{-1/\kappa})\over
       \Li_{\alpha-2}(\e^{-1/\kappa})-\Li_{\alpha-1}(\e^{-1/\kappa})},
\end{equation}
where $\Li_n(x)$ is the $n$th polylogarithm of~$x$.  Below this transition
no epidemics are possible, only small outbreaks having average size
\begin{equation}
\av{s} = 1 + {T[\Li_{\alpha-1}(\e^{-1/\kappa})]^2\over
          \Li_\alpha(\e^{-1/\kappa})
          [(T+1)\Li_{\alpha-1}(\e^{-1/\kappa})
           - T\Li_{\alpha-2}(\e^{-1/\kappa})]},
\end{equation}
while above it, epidemics occur with size and probability~$S$, whose value
we can extract by numerical iteration of Eq.~\eref{epidemic}.

\begin{figure}
\resizebox{11cm}{!}{\includegraphics{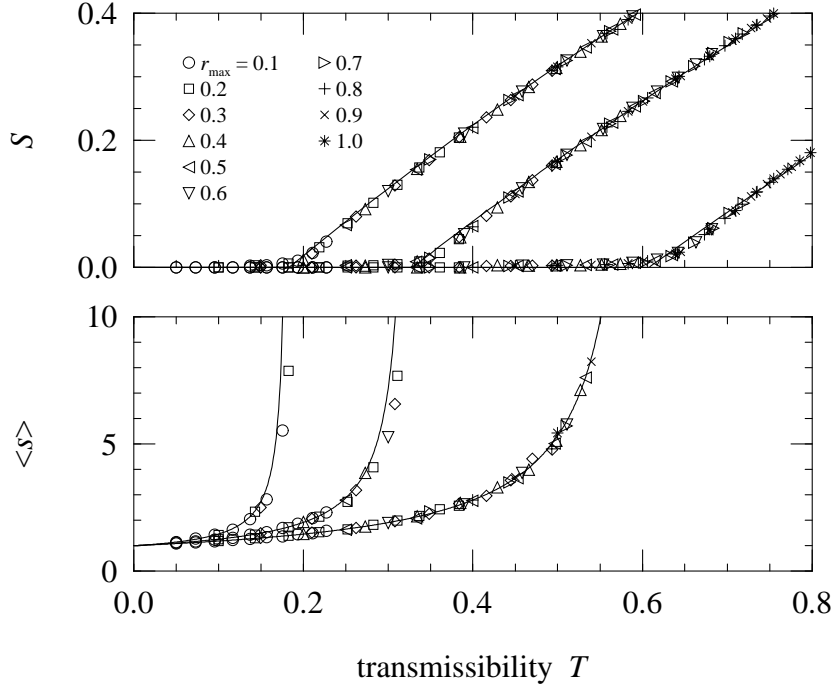}}
\caption{Epidemic size (top) and average outbreak size (bottom) for the SIR
  model on networks with degree distributions of the form
  Eq.~\eref{powerlaw} as a function of transmissibility.  Solid lines are
  the exact solutions, Eqs.~\eref{avs} and~\eref{epidemic}, for $\alpha=2$
  and (left to right in each panel) $\kappa=20$, $10$, and~$5$.  Each of
  the points is an average result for $10\,000$ simulations on graphs of
  $100\,000$ vertices each.  The distributions $P(r)$ and $P(\tau)$ are
  uniform over the intervals $0\le r<r_{\rm max}$ and $1\le\tau\le\tau_{\rm
    max}$ respectively ($r$~real, $\tau$~integer), with $r_{\rm max}$ as
  indicated and $\tau_{\rm max}=1\ldots10$.}
\label{siruni}
\end{figure}

In Fig.~\ref{siruni} we compare the predictions of this solution against
explicit simulations of epidemics spreading on networks with heterogeneous
transmission rates~$r$ and infectiveness times~$\tau$.  As the figure
shows, agreement between analytic and numerical results is good.

To emphasize the difference between our results and those for the
equivalent fully mixed model, we compare the position of the epidemic
threshold in the two cases.  In the case $\alpha=2$, $\kappa=10$ (the
middle curve in each frame of Fig.~\ref{siruni}), our analytic solution
predicts that the epidemic threshold occurs at $T_c=0.329$.  The
simulations agree well with this prediction, giving $T_c=0.32(2)$.  By
contrast, a fully mixed SIR model in which each infective individual
transmits the disease to the same average number of others as in our
network, gives a very different prediction of $T_c=0.558$.

Although the model above is already more realistic than the standard
epidemic models in a number of ways (network structure, heterogeneous
transmission, heterogeneous infectiveness times), there are many ways it
can be further improved.  For instance, with real diseases the transmission
rates~$r$ or the infectiveness times~$\tau$ may not be iid random variables
as we have assumed; they may be correlated.  As an example of how this can
be incorporated into the model, consider the case where the distribution of
transmission rates $r$ depends on the degree $k$ of the vertex representing
the infective individual.  (One could imagine for example that people who
have many contacts tend also to have more fleeting contacts, so that $r$
would go down on average with increasing~$k$.)  Then the transmissibility
also becomes a function of~$k$ according to $T_k = 1-\int \d r\>\d\tau \>
P_k(r) P(\tau)\, \e^{-r\tau}$ and the generating functions become a
function of the complete set $\set{T_k}$.  Alternatively, the distribution
of $r$ might depend on the degree of the individual \emph{being infected,}
which gives us a similar set $\set{U_k}$ of transmissibilities.  Or $r$
might depend on both degrees.  The correct generalization of the generating
functions is:
\begin{eqnarray}
\label{g0tkuk}
G_0(x;\set{T_k},\set{U_k}) &=& \sum_k p_k (1+(x-1)T_k)^k,\\
\label{g1tkuk}
G_1(x;\set{T_k},\set{U_k}) &=& 
    {1\over z} \sum_k kp_k [1+((1+(x-1)T_k)^{k-1}-1)U_k].
\end{eqnarray}
The cases in which transmission depends only on one degree or the other can
be derived from these expressions by setting either $T_k=1$ or $U_k=1$ for
all~$k$.  Once we have the generating functions, then the calculation
proceeds as before, with mean outbreak size below the epidemic transition
being given by Eq.~\eref{avs} and epidemic size above it by
Eq.~\eref{epidemic}.  The epidemic transition occurs as before at
$G_1'(1;\set{T_k},\set{U_k})=1$.

Another area of current interest is models incorporating vaccination of
individuals~\cite{MN00a,PV01c}.  We show elsewhere~\cite{CNSW00,Newman02b}
that models with vaccination can also be solved exactly, both in the case
of uniform independent vaccination probability (i.e.,~random vaccination of
a population) and in the case of vaccination which is correlated with
properties of individuals such as their degree (so that vaccination can be
directed at the so-called core group of the disease-carrying
network---those with the highest degrees).

The other main way in which we can make our models more realistic, while
still retaining exact solvability, is to incorporate more realistic social
structure into our networks.  As an example, consider the network by which
a sexually transmitted disease is communicated, which is also the network
of sexual partnerships between individuals.  In a recent study of 2810
respondents Liljeros~\etal~\cite{Liljeros01}\ recorded the numbers of
sexual partners of men and women over the course of a year.  From their
data it appears that the distributions of these numbers are power-law in
form $p_k\sim k^{-\alpha}$ for both men and women with exponents $\alpha$
that fall in the range $3.1$ to~$3.3$.  If we assume that the disease of
interest is transmitted primarily by contacts between men and women (true
only for some diseases), then to a good approximation the network of
contacts is bipartite~\cite{NSW01}.  We define two pairs of generating
functions for males and females:
\begin{eqnarray}
F_0(x) &=& \sum_j p_j x^j,\qquad
F_1(x) = \frac{1}{\mu} \sum_j j p_j x^{j-1},\\
G_0(x) &=& \sum_k q_k x^k,\hspace{7.8mm}
G_1(x) = \frac{1}{\nu} \sum_k k q_k x^{k-1},
\end{eqnarray}
where $p_j$ and $q_k$ are the two degree distributions and $\mu$ and $\nu$
are their means.  We can then develop expressions similar to
Eqs.~\eref{avs} and~\eref{epidemic} for an epidemic on this new network.
For instance, the epidemic transition takes place at the point where
$T_{mf} T_{fm} = 1/[F_1'(1)G_1'(1)]$ where $T_{mf}$ and $T_{fm}$ are the
transmissibilities for male-to-female and female-to-male infection
respectively.

One important result that follows immediately is that if the degree
distributions are truly power-law in form, then there exists an epidemic
transition only for a small range of values of the exponent $\alpha$ of the
power law.  Let us assume, as appears to be the case, that the exponents
are roughly equal for men and women: $\alpha_m=\alpha_f=\alpha$.  Then if
$\alpha\le3$, we find that $T_{mf} T_{fm} = 0$, which is only possible if
at least one of the transmissibilities $T_{mf}$ and $T_{fm}$ is zero.  As
long as both are positive, we will always be in the epidemic regime, and
this would clearly be bad news.  No amount of precautionary measures to
reduce the probability of transmission would ever eradicate the disease.
(Similar results have been seen in other types of models
also~\cite{PV01a,LM01}.)  Conversely, if $\alpha>\alpha_c$, where
$\alpha_c=3.4788\ldots$ is the solution of
$\zeta(\alpha-2)=2\zeta(\alpha-1)$, we find that $T_{mf} T_{fm}=1$, which
is only possible if both $T_{mf}$ and $T_{fm}$ are~1.  When either is less
than~1 no epidemic will ever occur, which would be good news.  Only in the
small intermediate region $3<\alpha<3.4788\ldots$ does the model possess an
epidemic transition.  Interestingly, the real-world network measured by
Liljeros~\etal~\cite{Liljeros01} appears to fall precisely in this region,
with $\alpha\simeq3.2$.  If true, this would be both good and bad news.  On
the bad side, it means that epidemics can occur.  But on the good side, it
means that that it is in theory possible to prevent an epidemic by reducing
the probability of transmission, which is precisely what most health
education campaigns attempt to do.  The predicted critical value of the
transmissibility is $\zeta(\alpha-1)/[\zeta(\alpha-2)-\zeta(\alpha-1)]$,
which gives $T_c=0.363\ldots$ for $\alpha=3.2$.  Epidemic behavior would
cease were it possible to arrange that $T_{mf}T_{fm}<T_c^2$.

Some caveats are in order here.  The error bars on the values of the
exponent $\alpha$ are quite large (about $\pm0.3$~\cite{Liljeros01}).
Thus, assuming that the conclusion of a power-law degree distribution is
correct in the first place, it is still possible that $\alpha<3$, putting
us in the regime where there is always epidemic behavior regardless of the
value of the transmissibility.  On the other hand, it may also be that the
distribution is not a perfect power law.  Although the measured
distributions do appear to have power-law tails, it seems likely that these
tails are cut off at some point.  If this is the case, then there will
always be an epidemic transition at finite~$T$, regardless of the value
of~$\alpha$.  Furthermore, if it were possible to reduce the number of
partners that the most active members of the network have, so that the
cutoff moves lower, then the epidemic threshold rises, making it easier to
eradicate the disease.  Interestingly, the fraction of individuals in the
network whose degree need change in order to make a significant difference
is quite small.  At $\alpha=3$, for instance, a change of cutoff from
$\kappa=\infty$ to $\kappa=100$ affects only 1.3\% of the population, but
increases the epidemic threshold from $T_c=0$ to $T_c=0.52$.  In other
words, targeting preventative efforts at changing the behavior of the most
active members of the network may be a much more promising way of
preventing the spread of disease than targeting everyone.  (This suggestion
is certainly not new, but our models provide a quantitative basis for
assessing its efficacy.)

Another application of the techniques presented here is described in
Ref.~\onlinecite{AN01}.  In that paper we model in detail the spread of
walking pneumonia ({\em Mycoplasma pneumoniae\/}) in a closed setting (a
hospital) for which network data are available from observation of an
actual outbreak.  In this example, our exact solutions agree well both with
simulations and with data from the outbreak studied.  Furthermore,
examination of the analytic solution allows us to make specific suggestions
about possible new control strategies for {\em M.  pneumoniae\/} infections
in settings of this type.

Applications of the techniques described here are also possible for
networks specific to many other settings, and hold promise for the better
understanding of the role that the structure of contact networks plays in
the spread of disease.

{\bigskip\small The author thanks Lauren Ancel, L\'aszl\'o Barab\'asi,
  Duncan Callaway, Michelle Girvan, and Catherine Macken for useful
  comments.  This work was supported in part by the National Science
  Foundation under grant number DMS--0109086.}

\medbreak

\end{document}